\documentclass[aps,pra,twocolumn,superscriptaddress,amsmath,amssymb,preprintnumbers,showpacs]{revtex4}

\usepackage{graphicx}
\usepackage{dcolumn}

\begin{document}


\title{Lindblad and non--Lindblad type dynamics of a quantum Brownian particle}


\author{S. Maniscalco}

\affiliation{School of Pure and Applied Physics, University of
KwaZulu-Natal, Durban 4041, South Africa}
\email{maniscalco@ukzn.ac.za}

\affiliation{INFM, MIUR and Dipartimento di Scienze Fisiche ed
Astronomiche dell'Universit\`{a} di Palermo, via Archirafi 36,
90123 Palermo, Italy}

\author{J.Piilo}
\affiliation{School of Pure and Applied Physics, University of
KwaZulu-Natal, Durban 4041, South Africa}

\affiliation{Department of Physics, University of Turku, FIN-20014
Turun yliopisto, Finland}

\affiliation{Helsinki Institute of Physics, PL 64, FIN-00014 Helsingin yliopisto, Finland}

\author{F. Intravaia}
\affiliation{Laboratoire Kastler Brossel \footnote{\'{E}cole
Normale Sup\'{e}rieure, Centre National de la Recherche
Scientifique, Universit\'{e} Pierre et Marie Curie.}, Case 74, 4
place Jussieu, F-75252 Paris Cedex 05, France}

\author{F. Petruccione}

\affiliation{School of Pure and Applied Physics, University of
KwaZulu-Natal, Durban 4041, South Africa}

\author{A. Messina}
\affiliation{INFM, MIUR and Dipartimento di Scienze Fisiche ed
Astronomiche dell'Universit\`{a} di Palermo, via Archirafi 36,
90123 Palermo, Italy}

\date{\today}

\begin{abstract}
The dynamics of a typical open quantum system, namely a quantum
Brownian particle in a harmonic potential, is studied focussing on
its non-Markovian regime. Both an analytic approach and a
stochastic wave function approach are used to describe the exact
time evolution of the system. The border between two very
different dynamical regimes, the Lindblad and non-Lindblad
regimes, is identified and  the relevant physical variables
governing the passage from one regime to the other are singled
out. The non-Markovian short time dynamics is studied in detail by
looking at the mean energy, the squeezing, the Mandel parameter
and the Wigner function of the system.

\end{abstract}


\maketitle

\section{Introduction}
The description of the dynamics of quantum systems interacting
with their surrounding is, in general, a very difficult task due
to the complexity of the environment. An exact approach would
require to take into account not only the degrees of freedom of
the system of interest, but also those of the environment. One
can, in principle, look for the solution of the Liouville - von
Neumann equation for the density matrix of the total closed
system. Even when this is possible, however, the total density
matrix contains much more information than what we actually need,
since one is usually interested in the time evolution of the
reduced system only.

A common approach to the dynamics of open quantum systems consists
in deriving a master equation for the reduced density matrix
describing the temporal behavior of the open system
\cite{petruccionebook}. This equation is in general obtained by
tracing over the environmental variables, after performing a
series of approximations. Two of the most common ones are the
Rotating Wave Approximation (RWA) and the Born-Markov
approximation. The first one basically consists in neglecting in
the microscopic system-reservoir interaction Hamiltonian the
counter-rotating terms responsible for the virtual exchanges of
energy between system and environment. The second one neglects the
correlations between system and reservoir assuming that the
changes in the reservoir due to the interaction with the system
cannot feed back into the system's dynamics.

The Born-Markov approximation leads to a master equation which can
be cast in the so called Lindblad form \cite{Lindblad,Gorini}.
Master equations in the Lindblad form are characterized by the
fact that the dynamical group of the system satisfies both the
semigroup property and the complete positivity condition, thus
ensuring the preservation of positivity of the density matrix
during the time evolution. Moreover, it has been shown that
numerical techniques as the Monte Carlo Wave Function method can
always, in principle, be applied to the description of the
dynamics, provided that the master equation is in the Lindblad
form \cite{Molmer96a}.

In many solid state systems such as photonic band gap materials
and quantum dots the Markov approximation is, however, not
justified \cite{quang97}. Similarly, the reservoir interacting
with a single mode cavity in atom lasers is strongly non-Markovian
\cite{hope00}. These physical systems therefore necessitate
non-Markovian analytical or numerical approaches to their
dynamics. Moreover, the non-Markovian features become of
importance when one is interested in the initial temporal regime,
even for Markovian systems, where the memory time of the reservoir
$\tau_R$ is much smaller than the system characteristic time scale
$\tau_S$.

It is worth mentioning that during the last few years the interest
in open quantum systems has increased mainly due to three reasons.
On the one hand, the phenomena of decoherence and dissipation,
characterizing the dynamics of a quantum system interacting with
its surrounding \cite{giulini}, are considered nowadays the major
obstacles to the realization of quantum computers and other
quantum devices \cite{nielsen}. On the other hand recent
experiments on engineering of environments \cite{engineerNIST}
pave the way to new proposals aimed at creating entanglement and
superpositions of quantum states exploiting decoherence and
dissipation \cite{cirac,plenio}. Moreover, we emphasize that
presently the fields of quantum information and open quantum
systems actually merge in the question to what extent the
entangling power of the system-reservoir interaction is the
responsible factor for decoherence \cite{plenio2,dodd}. In order
to understand deeply the origin, essence and effects of
decoherence phenomena, it is of great importance to have the tools
for an exact description of the open system dynamics.

In this paper we focus on an ubiquitous model of the theory of
open quantum systems, that is the quantum Brownian motion model,
which is also known as the damped harmonic oscillator
\cite{petruccionebook,Haake,annals,Feynman,Caldeira,Hu,Ford,Grabert,PRAsolanalitica,letteranostra,EPJRWA,misbelief}.
We consider the exact generalized master equation obtained with
the time-convolutionless projection operator technique
\cite{petruccionebook}. We exploit an analytic approach based on
the algebraic properties of the superoperators appearing in the
master equation \cite{PRAsolanalitica}. The knowledge of the
explicit expression of the exact analytic solution of this
equation allows us to study in a complete way the dynamics and, in
particular, the short time non-Markovian regime. The analytic
results are in very good agreement with the numerical simulations
obtained by means of the non-Markovian  Wave Function (NMWF)
method \cite{Breuer99a}, a variant of the Monte Carlo (MC) methods
\cite{Dalibard92a,Molmer93a,Molmer96a,Plenio98a,Breuer04a}. By
looking at the dissipation and diffusion coefficients we single
out those parameters governing the passage from Lindblad to
non-Lindblad type regimes. The dynamics of the system in the
Lindblad type regime is governed by a Lindblad type master
equation, whereas in the non-Lindblad type regime, the master
equation cannot be cast in the Lindblad form and the dynamics is
dominated by virtual exchanges of energy between the system and
the environment.

The paper is structured as follows. In Sec. \ref{sec:exact} we
recall the exact time-convolutionless master equation for quantum
Brownian motion and its superoperatorial solution. In Sec.
\ref{sec:MC} we apply the non-Markovian Wave Function method to
the system under scrutiny. Sections \ref{sec:results} and
\ref{sec:results2} contain the main results of the paper. In Sec.
\ref{sec:results} we study the border between Lindblad and
non-Lindblad regions, and in Sec. \ref{sec:results2} we focus on
the non-Lindblad type dynamics looking at the temporal evolution
of the squeezing, of the Mandel parameter and of the Wigner
function. Finally,
 Sec. \ref{sec:conclusions} concludes the paper.

\section{Exact dynamics of quantum Brownian particle}
\label{sec:exact}

In this section we first recall the microscopic model and the
exact master equation for the reduced density matrix of a quantum
Brownian particle. Then we briefly sketch an analytic approach to
derive an exact solution for the master equation.

Let us consider a harmonic oscillator of frequency $\omega_0$
surrounded by a generic bosonic environment. The Hamiltonian $H$
of the total system can be written as follows
\begin{equation}
H=H_0 + H_E + \alpha X E, \label{eq:secI1}
\end{equation}
where $H_0=\omega_0 \left(P^2+X^2\right)/2$, $H_E$ and $\alpha X
E$ are the system, environment and interaction Hamiltonians,
respectively, and $\alpha$ is the dimensionless coupling constant.
The interaction Hamiltonian considered here has a simple bilinear
form with position of the oscillator $X$ and position
environmental operator $E \equiv \sum_n \kappa_n x_n$, where $x_n$
are the position operators of the environmental oscillators.  For
the sake of simplicity we have written the previous expressions in
terms of dimensionless position and momentum operators for the
system oscillator. The key quantity governing the nature of the
coupling is the spectral density $J(\omega) = \sum_n \kappa_n
\delta (\omega-\omega_n)/(2 m_n \omega_n)$, with $m_n$ and
$\omega_n$ masses and frequencies of the environmental
oscillators, respectively.

We denote by $ \rho$ the density matrix of the total system. Under
the  assumptions that: (i)  at $t=0$ system and environment are
uncorrelated, that is $\rho(0)= \rho_S (0) \otimes \rho_E(0)$,
with $\rho_S$ and $ \rho_E$ density matrices of the system and the
environment respectively; (ii) the environment is stationary, that
is $[H_E,\rho_E(0)]=0$; (iii) the expectation value of the
environmental operator $E$  is zero, that is ${\rm Tr_E} \left\{ E
\rho_E(0) \right\}=0$ (as for example in the case of a thermal
reservoir), one can derive the following master equation
\cite{EPJRWA}
\begin{eqnarray}
\frac{d \rho_S(t)}{dt} = \frac{1}{i \hbar} \mathbf{H}_{0}^S
\rho_S(t)
 -  \Big[ \Delta(t) (\mathbf{X}^S)^2 - \Pi(t) \mathbf{X}^S
\mathbf{P}^S \nonumber \\
 - \frac{i}{2} r(t) (\mathbf{X^2})^S + i \gamma(t) \mathbf{X}^S
\mathbf{P}^{\Sigma} \Big]\rho_S(t). \label{QBMme}
\end{eqnarray}
We indicate with $\mathbf{X}^{S(\Sigma)}$ and
$\mathbf{P}^{S(\Sigma)}$ the commutator (anticommutator) position
and momentum superoperators respectively and with
$\mathbf{H}_{0}^S$ the commutator superoperator relative to the
system Hamiltonian. It is not difficult to prove that such a
master equation, obtained by using the time-convolutionless
projection operator technique
\cite{petruccionebook,annals,timeconv}, is the superoperatorial
version of the Hu-Paz-Zhang master equation \cite{Hu}. This
equation can be solved in generality
\cite{Haake,Hu,Ford,Grabert,PRAsolanalitica} although the time
dependent coefficients have no obvious closed form. For this
reason, the exact study of the non-Markovian behaviour of the
system for strong couplings may be performed only by means of
numerical methods. Also in this case, however, one needs, firstly,
to find out the parameter regime in which a perturbation expansion
to a given order yields reliable numerical results. This is in
general a very difficult task. In the following we will focus on
the dynamics of the system in the weak coupling limit, and in
particular we will consider the case in which a truncation of the
expansion in the coupling strength $\alpha$ to the second order is
physically meaningful. Under this condition the time dependent
coefficients appearing in the master equation can be written as
follows
\begin{eqnarray}
\Delta(t)&=& \int_0^t \kappa(\tau) \cos (\omega_0 \tau) d \tau,
\label{delta} \\
\gamma(t) &=& \int_0^t \mu(\tau) \sin (\omega_0
\tau) d \tau, \label{gamma} \\
\Pi(t) &=& \int_0^t \kappa(\tau) \sin (\omega_0
\tau) d \tau, \label{pi} \\
r(t)&=& 2 \int_0^t \mu(\tau) \cos (\omega_0 \tau) d \tau
\label{erre},
\end{eqnarray}
where
\begin{equation}
\kappa(\tau)= \alpha^2 \langle \{ E(\tau),E(0)\} \rangle,
\label{kappa}
\end{equation}
and
\begin{equation}
\mu(\tau)= i \alpha^2 \langle [ E(\tau),E(0)] \rangle, \label{mu}
\end{equation}
are the noise and dissipation kernels, respectively. For the case
of an Ohmic reservoir spectral density with Lorentz-Drude cutoff
\cite{petruccionebook}
\begin{equation}
J(\omega)= \frac{2  \omega}{\pi} \
\frac{\omega_c^2}{\omega_c^2+\omega^2}, \label{spectrald}
\end{equation}
the noise and dissipation kernels assume the form
\begin{eqnarray}
\kappa(\tau) &=& 4 \alpha^2 k T \omega_c^2
\sum_{n=-\infty}^{\infty} \frac{\omega_c e^{-\omega_c \tau }-\nu_n
e^{-|\nu_n|\tau}}{\omega_c^2-\nu_n^2}, \label{kappa2} \\
\mu(\tau) &=& 2 \alpha^2 \omega_c^2 e^{-\omega_c \tau
},\label{mu2}
\end{eqnarray}
where $\omega_c$ is the cutoff frequency and $\nu_n=2 \pi n kT$
denote the Matsubara frequencies.

It is worth noting that, as shown in Ref. \cite{annals}, it is
possible to estimate in an easy way the order of magnitude of the
error associated to the truncated expression of the coefficients.
This allows to check  the range of validity of the weak coupling
approximation. The errors of the time dependent coefficients, up
to the fourth order in the coupling constant, are studied in Refs.
\cite{petruccionebook,annals}.

As we will see in the following, truncating the perturbation
expansion to the second order, it is possible to find a closed
analytic form for two time dependent coefficients playing a
special role in the dynamics: the diffusion coefficient
$\Delta(t)$ and the dissipation coefficient $\gamma(t)$. Dealing
with a closed analytic expression of these parameters allows to
gain new insight in the dynamics of the open system. In fact, the
possibility of studying analytically the border between Lindblad
and non-Lindblad-type dynamics stems from the availability of a
closed expression for these time dependent parameters.

Let us now look in more detail at the form of the master equation
(\ref{QBMme}). First of all, we note that this master equation is
local in time, even if non-Markovian. This feature is typical of
all the generalized master equations derived by using the
time-convolutionless projection operator technique
\cite{petruccionebook,Breuer99a} or equivalent approaches such as
the superoperatorial  one \cite{EPJRWA,royer}.

The time dependent coefficients appearing in (\ref{QBMme}) contain
all the information about the short time system-reservoir
correlation. The coefficient $r(t)$ gives rise to a time dependent
renormalization of the frequency of the oscillator. In the weak
coupling limit one can show that $r(t)$ gives a negligible
contribution as far as the reservoir cutoff frequency remains
finite \cite{petruccionebook}. The term proportional to
$\gamma(t)$ is a classical damping term while the coefficients
$\Delta(t)$ and $\Pi(t)$ are diffusive terms.
Averaging over the rapidly oscillating terms appearing in the time
dependent coefficients of Eq.(\ref{QBMme}) one gets the following
secular approximated master equation
\begin{eqnarray}
\frac{ d \rho_S}{d t}= \!\!&-& \!\!\! \frac{\Delta(t) + \gamma
(t)}{2} \left[ a^{\dag} a \rho_S - 2 a \rho_S a^{\dag} + \rho_S
a^{\dag} a \right]
\nonumber \\
\!\! &-& \!\!\!\frac{\Delta(t) - \gamma (t)}{2} \left[  a a^{\dag}
\rho_S - 2 a^{\dag} \rho_S a + \rho_S a a^{\dag}
 \right]\!\!, \label{MERWA}
\end{eqnarray}
where we have introduced the bosonic annihilation and creation
operators $a=\left( X+i P \right)/\sqrt{2}$ and $a^{\dag}=\left(
X-i P \right)/\sqrt{2}$. The form of Eq. (\ref{MERWA}) is similar
to the Lindblad form, with the only difference that the
coefficients appearing in the master equation are time dependent.
We say that this master equation is of Lindblad type when the
coefficients $\Delta(t) \pm \gamma (t)$ are positive at all times
\cite{misbelief}. Note, however, that Lindblad type master
equations, contrarily to master equations of Lindblad form, in
general do not satisfy the semigroup property.

In what follows we focus on the secular master equation given by
Eq. (\ref{MERWA}). Let us stress that the secular approximation
invoked here  does not coincide with the RWA commonly used to
describe quantum optical systems. Indeed, as shown in Ref.
\cite{EPJRWA}, differences in what we may call the {\it RWA
performed before or after tracing over the environment} do exist,
and they are in principle measurable. The {RWA performed before
tracing over the environment} consists in neglecting the
counter--rotating terms in the microscopic Hamiltonian describing
the coupling between system and environment. The {RWA performed
after tracing over the environment} is more precisely a secular
approximation, consisting in an average over rapidly oscillating
terms, but does not wash out the effect of the counter--rotating
terms present in the coupling Hamiltonian (see also Ref.
\cite{misbelief}).

It is worth noting that there exists a class of observables not
influenced by the secular approximation
\cite{PRAsolanalitica,grab}. The exact time evolution of the
operators belonging to such a class can be obtained by solving Eq.
(\ref{MERWA}). Examples of such observables are the mean value of
the quantum number operator $\langle n (t) \rangle$, hereafter
called the heating function, and the Mandel parameter $Q$.

For the Ohmic spectral density introduced in
Eq.~(\ref{spectrald}), the analytic expression for the dissipation
coefficient $\gamma(t)$, to second order in the coupling constant
is
\begin{equation}
\gamma (t)\! \!=\!\! \frac{\alpha^2 \omega_0 r^2}{r^2+1} \Big[1
\!-\! e^{- \omega_c t} \cos(\omega_0 t) \! - r e^{- \omega_c t}
\sin( \omega_0 t )  \Big], \label{gammasecord}
\end{equation}
with $r=\omega_c/\omega_0$.

As for the diffusion coefficient $\Delta(t)$, defined in Eq.
(\ref{delta}), a simple analytic expression is obtained only in
the high and low temperature regimes. In Appendix A we give the
expression for $\Delta(t)$ for generic $T$ and in Appendix B its
high $T$ and Markovian approximations.


The master equation (\ref{QBMme}) can be exactly solved by using
specific algebraic properties of the superoperators
\cite{PRAsolanalitica}. The solution for the density matrix of the
system is derived in terms of the quantum characteristic function
(QCF) $\chi_t(x,p)$ at time $t$, defined through the relation
\cite{cohen}
\begin{equation}
\label{sdef} \rho_S(t)=\frac{1}{2\pi}\int \chi_t(x,p)\:
e^{-i\left(p X-x P\right)} dxdp.
\end{equation}
It is worth noting that one of the advantages of this approach is
the relatively easiness in calculating the analytic expressions
for the mean values of observables of interest by means of the
relations:
\begin{eqnarray}
\langle X^n \rangle &=& (-i)^n\left(\frac{\partial^n}{\partial p^n
}\chi(x,p)\right)_{x,p=0},\nonumber \\
\label{xp}\langle P^n \rangle
&=&(i)^n\left(\frac{\partial^n}{\partial x^n
}\chi(x,p)\right)_{x,p=0}.
\end{eqnarray}
In the secular approximation the QCF is found to be
\cite{PRAsolanalitica}
\begin{equation}
\chi_t (x,p)=e^{- \Delta_{\Gamma}(t) (x^2+p^2)/2} \chi_0 \left(
e^{- \Gamma (t)} \tilde{x}, e^{- \Gamma (t)} \tilde{p}  \right),
\label{chit}
\end{equation}
where $\chi_0$ is the QCF of the initial state of the system, and
we defined
\begin{eqnarray}
\tilde{x} &=& \cos(\omega_0 t) x + \sin(\omega_0 t) p, \nonumber \\
\tilde{p} &=& - \sin(\omega_0 t) x + \cos(\omega_0 t) p.
\end{eqnarray}
 The quantities $\Delta_{\Gamma}(t)$ and $\Gamma(t)$ appearing in Eq.
 (\ref{chit})
 are defined in terms of the diffusion and dissipation
 coefficients $\Delta(t)$ and $\gamma(t)$ respectively (see
 Eqs. (\ref{gammasecord}) and  (\ref{deltasecord})) as follows
\begin{eqnarray}
\Gamma(t)&=& 2\int_0^t \gamma(t_1)\:dt_1, \label{Gamma} \\
\Delta_{\Gamma}(t) &=& e^{-\Gamma(t)}\int_0^t
e^{\Gamma(t_1)}\Delta(t_1)dt_1 \label{DeltaGamma}.
\end{eqnarray}
Equation~(\ref{chit}) shows that the QCF is the product of an
exponential factor, depending on both the diffusion $\Delta(t)$
and the dissipation $\gamma(t)$ coefficients, and a transformed
initial QCF. The exponential term is responsible of energy
dissipation and it is independent of the initial state of the
system. Information on the initial state is given by the second
term of the product, the transformed initial QCF. In the weak
coupling limit considered here the asymptotic values of the
diffusion and dissipation coefficients coincide with the Markovian
ones (see Appendix B). In this case $ \chi_0 \left[ e^{- \Gamma
(t)} \tilde{x}, e^{- \Gamma (t)} \tilde{p} \right] \rightarrow 1$
for long times, and the system approaches, as one would expect, a
thermal state at reservoir temperature, whatever the initial state
was. In general, however, for strongly coupled systems, the steady
state could be very different from the thermal state. For example,
in Ref. \cite{petruccionebook} (see pp. 481-483) it is shown that,
already for $\alpha / \omega_0 = 0.25$ the steady state solution
in the low temperature regime shows squeezing in position.

\section{Non-Markovian Wave Function Simulations}\label{sec:MC}

In this section we describe how to implement the non-Markovian
wave function (NMWF) method for the study of quantum Brownian
motion by using the stochastic unravelling of the master equation
in the doubled Hilbert space~\cite{Breuer99a,petruccionebook}. We
use Monte Carlo (MC)  methods both to confirm the validity of the
involved analytical solution and to demonstrate that these methods
can be used to study the heating dynamics of a quantum Brownian
particle in very general conditions. One might think that it is
straightforward to apply MC methods, e.g. the NMWF method, once
the master equation of the system and the corresponding  jump
operators are known. However, there exist situations in which the
MC simulations become exceedingly heavy from the computer resource
point of view~\cite{Piilo02a,Piilo03a}. In the following we show
that, in our case, MC methods can be used conveniently to study
numerically the system dynamics also in the non-Lindblad regime
where the time dependent decay coefficients $\Delta(t) \pm
\gamma(t)$ may acquire temporarily negative values.

\subsection{General form of non-Markovian wave function method in the doubled Hilbert space}\label{sec:doubled}
The most general form of the master equation obtained from
time-convolutionless projection operator technique
reads~\cite{petruccionebook,Breuer99a}
\begin{eqnarray}
\frac{\partial}{\partial t} \rho\left( t \right) &=& A \left( t
\right)  \rho \left( t \right) +  \rho \left( t \right)
B^{\dag}\left( t \right)
\nonumber\\
 &+&
 \sum_i C_i\left( t \right) \rho \left( t \right) D^{\dag}_i\left( t \right),
\label{eq:genmaster}
\end{eqnarray}
with time-dependent linear operators $A\left( t \right)$, $B\left(
t \right)$, $C_i\left( t \right)$, and $D_i\left( t \right)$.  The
unravelling of the master equation can be implemented by using the
method of stochastic unravelling in the doubled Hilbert space
\cite{petruccionebook} $\widetilde{\cal H}={\cal H}_S\oplus{\cal
H}_S$, where the state of the system is described by a pair of
stochastic state vectors
\begin{equation}
\theta\left( t \right) = \left(\begin{array}{c}
      \phi  \left( t \right) \\
      \psi\left( t \right)
\end{array} \right).
\end{equation}

The time-evolution of $\theta\left( t \right)$ can be described as
a piecewise deterministic process (PDP) \cite{petruccionebook}.
The deterministic part of the PDP is obtained by solving the
following differential equation
\begin{eqnarray}
\frac{\partial}{\partial t} \theta \left( t \right) = \left[
F\left( t \right)+\frac{1}{2}\sum_i \frac{\| J_i\left( t
\right)\theta\left( t \right) \|^2}{\| \theta\left( t \right)
\|^2} \right] \theta\left( t \right), \label{eq:time}
\end{eqnarray}
with
\begin{equation}
F\left( t \right) = \left(
\begin{array}{cc}
      A\left( t \right) & 0 \\
       0 &  B\left( t \right)
\end{array} \right)
\label{eq:f}
\end{equation}
and
\begin{equation}
J_i\left( t \right) =  \left( \begin{array}{cc}
      C_i\left( t \right) & 0 \\
      0                    & D_i\left( t \right)
\end{array} \right),\label{eq:ji}
\end{equation}
where $A\left( t \right)$, $B\left( t \right)$, $C_i\left( t
\right)$, and $D_i\left( t \right)$ are the operators appearing in
Eq.~(\ref{eq:genmaster}).

The stochastic part of the PDP is described in terms of jumps
inducing transitions of the form
\begin{equation}
\theta \left( t \right) \rightarrow \frac{\| \theta \left( t
\right) \|} {\| J_i\left( t \right)\theta \left( t \right) \|}
\left(
\begin{array}{c}
      C_i\left( t \right) \phi \left( t \right) \\
      D_i\left( t \right) \psi \left( t \right)
\end{array} \right).
\end{equation}
The jump rate for channel $i$ is given by
\begin{equation}
P_i (t)= \frac{\| J_i\left( t \right)\theta \left( t \right)
\|^2}{\| \theta \left( t \right) \|^2}.
\end{equation}
Finally, the solution for the reduced density matrix is obtained
as
\begin{equation}
\rho(t)=\int D \theta D \theta^{\ast} |\phi\rangle \langle\psi |
\tilde{P}[\theta,t],
\end{equation}
where $\tilde{P}[\theta,t]$ denotes the probability density
functional and the integration is carried out over the doubled
Hilbert space $\widetilde{{\cal
H}}$~\cite{petruccionebook,Breuer99a}.

\subsection{Implementation of the method for QBM}

The doubled Hilbert space state vector for the quantum Brownian
particle reads
\begin{equation}
\theta(t) = \left(
 \begin{array}{c}
      \phi  \left( t \right) \\
      \psi\left( t \right)
\end{array} \right)
= \left(
\begin{array}{c}
\sum_{n=0}^{\infty} \phi_n(t) |n\rangle \\
 \sum_{n=0}^{\infty} \psi_n(t) |n\rangle \\
\end{array} \right),
\end{equation}
where  $\phi_n(t)$ and $\psi_n(t)$ are the probability amplitudes
in the Fock state basis.

By comparing Eq.~(\ref{eq:genmaster}) with the master equation
(\ref{MERWA}), the operators $A(t)$ and $B(t)$ in Eq.~(\ref{eq:f})
have to be chosen as
\begin{eqnarray}
A(t) &=& B(t)= -i \omega_0 a^{\dag} a - \frac{1}{2} \left\{
\left[\Delta(t)+\gamma(t)\right]a^{\dag} a+ \right.
\nonumber \\
&& \left. \left[\Delta(t)-\gamma(t)\right]aa^{\dag} \right\}.
\end{eqnarray}
Accordingly, the operators $C_i$ and $D_i$ are
\begin{eqnarray}
C_1(t)&=&D_1(t)= \sqrt {|\Delta(t)-\gamma(t)|} a^{\dag},
\nonumber \\
C_2(t)&=&D_2(t)= \sqrt{|\Delta(t)+\gamma(t)|} a
\end{eqnarray}
and the corresponding operators $J_i$, given by Eq.(\ref{eq:ji})
become
\begin{eqnarray}
J_1(t)= \sqrt {|\Delta(t)-\gamma(t)|} \left(
\begin{array}{cc}
 {\rm sgn}\left[\Delta(t)-\gamma(t)\right] a^{\dag}& 0 \\
0& a^{\dag}\\
\end{array} \right)
\nonumber \\
J_2(t)= \sqrt {|\Delta(t)+\gamma(t)|} \left(
\begin{array}{cc}
 {\rm sgn}\left[\Delta(t)+\gamma(t)\right] a& 0 \\
0& a\\
\end{array} \right).
\end{eqnarray}
When the system dynamics and occupation of the states is
restricted to the two lowest Fock states the equations resemble
closely the ones used for the study of Jaynes-Cummings model with
detuning \cite{Breuer99a}.

The statistics of the quantum jumps is described by the waiting
time distribution function $F_w(\tau)$ which represents the
probability that the next jump occurs within the time interval
$[t,t+\tau)$. $F_w(\tau)$, derived from the properties of the
stochastic process, reads
\begin{equation}
F_w(\tau)=1-\exp\left[-\int_0^{\tau} \sum_{i=1,2}
P_i\left(s\right)ds\right],\label{eq:fw}
\end{equation}
where for channel 1 (jump up, the system absorbs a quantum of
energy from the environment)
\begin{equation}
P_1(t)=\frac{|\Delta(t)-\gamma(t)|}{\| \theta \left( t \right)
\|^2} \sum_{n=0}^{\infty} (n+1)\left[ |\phi_n(t)|^2  +
|\psi_n(t)|^2\right], \label{p1}
\end{equation}
and for channel $2$  (jump down, the system emits a quantum of energy
into the environment)
\begin{equation}
P_2(t)=\frac{|\Delta(t)+\gamma(t)|}{\| \theta \left( t \right)
\|^2} \sum_{n=0}^{\infty} n \left[ |\phi_n(t)|^2  +
|\psi_n(t)|^2\right]. \label{p2}
\end{equation}
When the jump occurs, the choice of the decay channel is made
according to the factors $P_1(t)$ and $P_2(t)$. The times at which
the jumps occur are obtained from Eq. (\ref{eq:fw}) by using the
method of inversion \cite{petruccionebook}.

\begin{figure*}[!htb]
\includegraphics{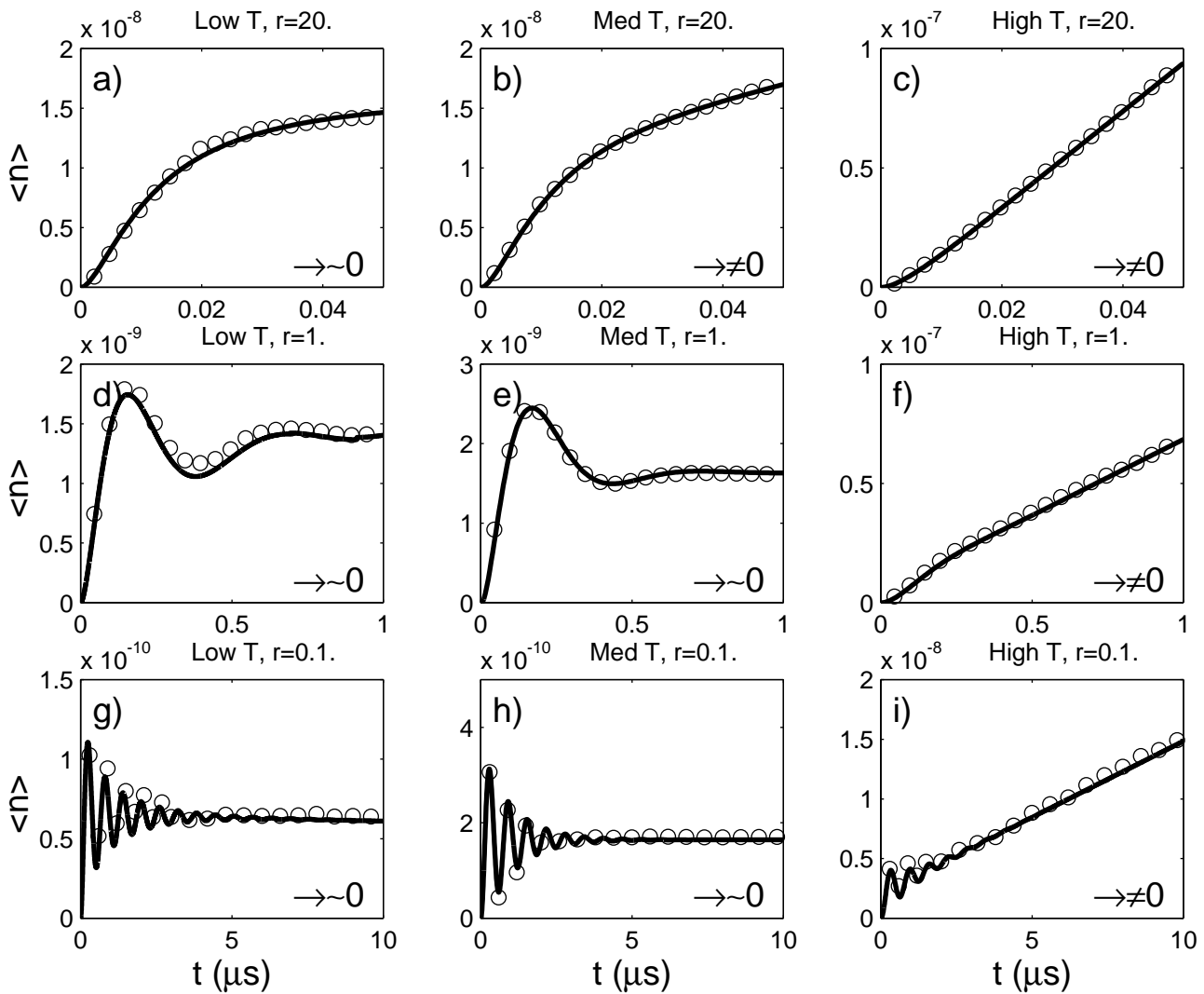}
\caption{\label{results1} Dynamics of the heating function
$\langle n(t) \rangle$ in the short time non-Markovian regime. For
the high $T$, graphics (c),(f),(i), we have used $r_0= \omega_0/
KT = 0.1$; for the intermediate $T$, graphics (b),(e),(h) we have
used $r_0= \omega_0/ KT = 1$; for low $T$, graphics (a),(d),(g),
we have used $r_0= \omega_0/ KT = 100$. We indicate with solid
line the analytic solution and with circles the simulations
performed using the NMWF method. In the right-lower corner of all
graphics we indicate whether the asymptotic long time value of the
heating function is null (zero $T$ reservoir) or not. }
\end{figure*}

For very low temperatures, the non-Markovian behavior of the
heating function of the quantum Brownian particle may occur when
$\langle n \rangle$ is of the order of $10^{-10}$, see
Fig.~\ref{results1}.  To reach such an accuracy, a MC simulation
for the estimation of $\langle n \rangle$  would require more than
$10^{10}$ realizations to be generated. This problem may be
circumvented by an appropriate scaling of the time dependent
coefficients $\Delta\left( t \right) \pm \gamma\left( t \right)$
of the master equation. The method is based upon the following
considerations. Let us look at  the properties of the Hilbert
space path integral solution of the stochastic process
corresponding to the unravelling of Eq.~(\ref{eq:genmaster}). The
Hilbert space path integral representation is essentially the
expansion of the propagator of the stochastic process
$T\left[\theta,t | \theta_0, t_0 \right]$ in the number of quantum
jumps \cite{petruccionebook}:
\begin{equation}
T\left[\theta,t | \theta_0, t_0 \right] =
\sum_{N=0}^{\infty}T^{(N)}\left[\theta,t | \theta_0, t_0 \right],
\end{equation}
where $N$ denotes the number of jumps, and $T^{(N)}$ are the $N$
jump contributions to the propagator. As long as in the time
evolution period of interest there is maximally one jump per
realization, it can be shown that, in the weak coupling limit and
for the initial conditions used here, the relevant contribution to
the propagator is given  by the first two terms $T^{(0)}$,
$T^{(1)}$. In this case the expectation value of an arbitrary
operator $O$ is given by
\begin{eqnarray}
\langle O \rangle \left( t \right) & =&
\int D \theta D \theta^* \langle \phi(t) | O | \psi(t)\rangle \nonumber \\
&& \left\{ T^{(0)}\left[\theta,t | \theta_0, t_0 \right]+
 T^{(1)}\left[\theta,t | \theta_0, t_0 \right]\right\}.
\end{eqnarray}
The contribution from $T^{(0)}$ gives the initial expectation
value $\langle O \rangle \left( 0 \right)$ plus a term which is
directly proportional to the decay coefficients $\Delta\pm\gamma$.
Since $T^{(1)}$ is also directly proportional to the decay
coefficients we get as a result that the change in the expectation
value is also proportional to the decay coefficients
\begin{equation}
\langle O \rangle \left( t \right) -\langle O \rangle \left( 0
\right)
 \propto
\Delta\pm\gamma.
 \end{equation}

Thus, to ease the numerics and still to obtain the correct result,
it is possible to speed up the decay by multiplying the
coefficients $\Delta\pm\gamma$ with some suitable factor $\beta$,
and to do the corresponding scaling down by dividing the
calculated ensemble average by the same factor at the end of the
simulation. For the heating function the validity of the scaling
can be seen directly from the analytic solution (see
Eq.(\ref{hfgenericT}) of the following section). The scaling
allows to reduce the ensemble size for the estimation of the
heating function from the unpractical $10^{10}$ to the more
practical $10^4-10^5$.

\section{The Lindblad - non--Lindblad border}\label{sec:results}

\begin{figure*}
\includegraphics[width=7.5 cm,height=7.5 cm]{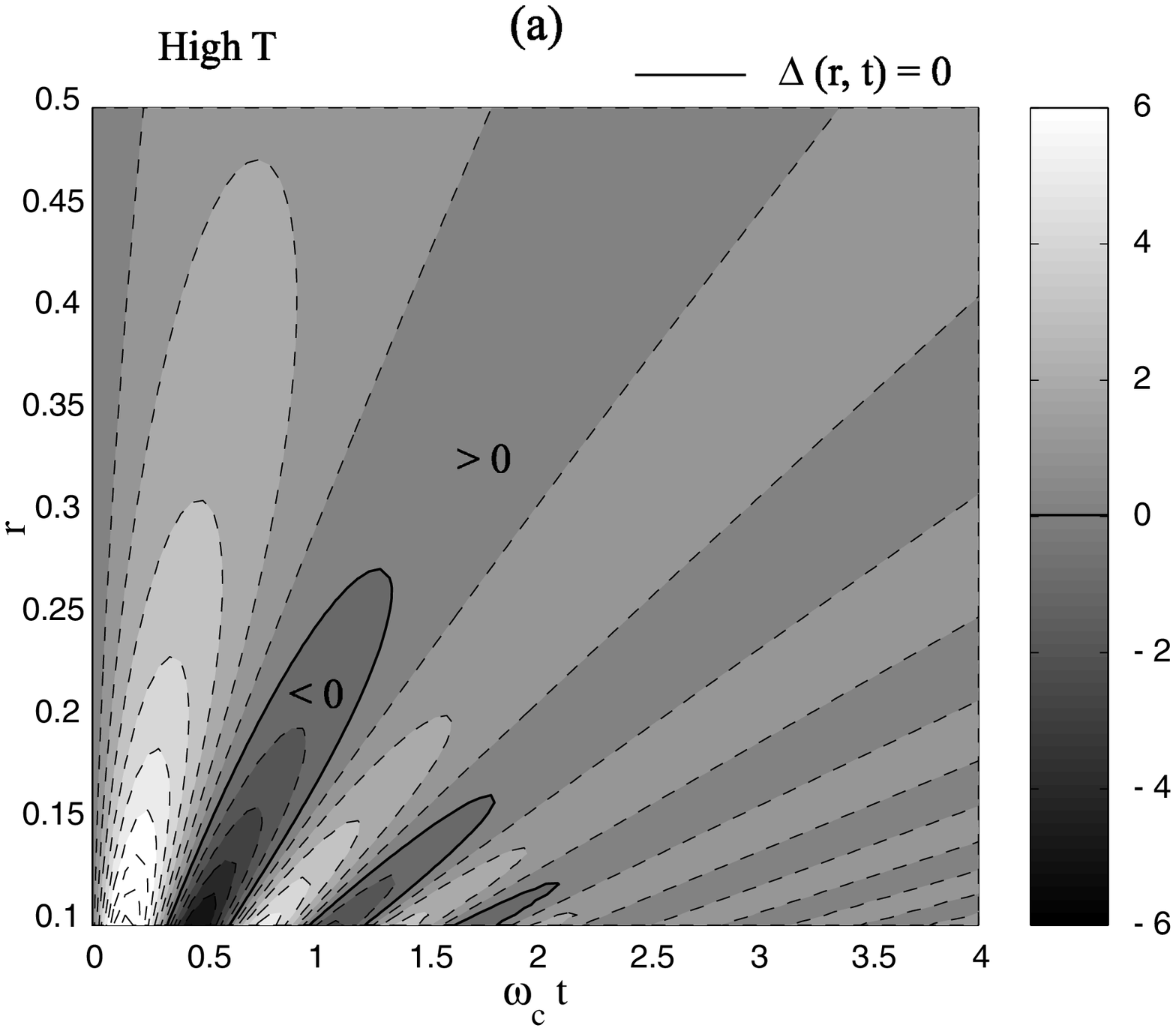}
\hspace{0.5cm}
\includegraphics[width=7.5 cm,height=7.5 cm]{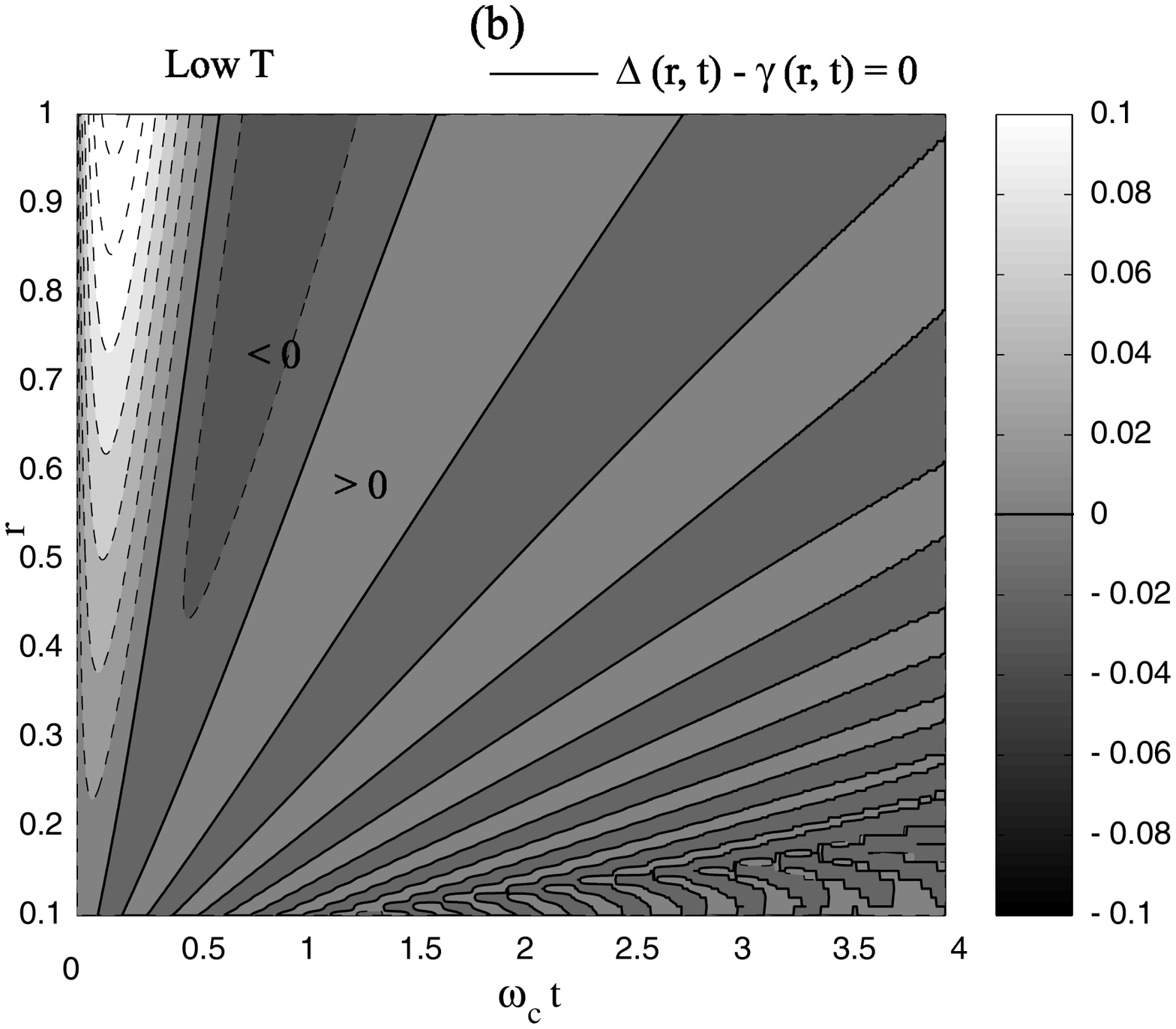}
\caption{\label{results2} Contour plots of $\overline{\Delta
(r,t)}=\Delta (r,t) / 2 \alpha^2 K T$ for high $T$ (a), and of the
coefficient $\overline{\Delta(r,t)-\gamma(r,t)} =
[\Delta(r,t)-\gamma(r,t)] / \alpha^2  $ for low $T$ (b). In (b) we
have chosen $2 \pi r_c = \hbar \omega_c / KT =10$. In both (a) and
(b) the contour line corresponding to $\overline{\Delta(r,t)}=0$
(a) and $\overline{\Delta(r,t)-\gamma(r,t)} =0$ is indicated by a
thick solid line.}
\end{figure*}

This section contains the main results of the paper. Stimulated by
the recent achievement in reservoir engineering techniques, we
look at the dynamics of a quantum Brownian particle for different
classes of reservoirs. We single out two reservoir parameters
playing a key role in the dynamics of the open system, i.e. its
temperature $T$ and the frequency cutoff $\omega_c$ of its
spectral density. As we will see in this section, by varying these
two parameters the time evolution of the system oscillator varies
from Lindblad type to non-Lindblad type.

\subsection{Heating function}
In order to illustrate the changes in the dynamics of the system
we will focus, first of all, on the temporal behaviour of the
heating function $\langle n(t) \rangle$. In the following section
we will further investigate the non-Markovian time evolution by
looking at the Mandel parameter, at the squeezing properties and
at the Wigner function of the system.

Having in mind Eq.~(\ref{chit}) and using Eq. (\ref{xp}), one gets
the following expression for the heating function
\begin{equation}
\label{hf} \langle n(t) \rangle = e^{-\Gamma(t)} \langle n(0)
\rangle + \frac{1}{2} \left(  e^{-\Gamma(t)}  - 1\right) +
\Delta_{\Gamma}(t),
\end{equation}
with $\Delta_{\Gamma}(t)$ and $\Gamma(t)$ defined by Eqs.
(\ref{Gamma}) and (\ref{DeltaGamma}).

The asymptotic long time behavior of the heating function, for
times $t$ much bigger than the reservoir correlation time $\tau_R
= 1/\omega_c$, is readily obtained by using the Markovian
stationary values for $\Delta(t)$ and $\gamma(t)$, as given by
Eqs. (\ref{deltaM}) - (\ref{gammaM}),
\begin{equation}
\langle n (t) \rangle =e^{- \Gamma t} \langle n(0) \rangle +
n(\omega_0) \left( 1- e^{- \Gamma t}\right), \label{nM}
\end{equation}
with $n(\omega_0)= \left( e^{ \hbar \omega_0/ K T
}-1\right)^{-1}$. This equation gives evidence for a second
characteristic time of the dynamics, namely the thermalization
time $\tau_T= 1/ \Gamma $, with $\Gamma= \alpha^2 \omega_0 r^2
/(r^2+1)$. The thermalization time depends both on the coupling
strength and on the ratio $r=\omega_c / \omega_0$ between the
reservoir cutoff frequency and the system oscillator frequency.
Usually, when studying QBM, one assumes that $r \gg 1$,
corresponding to a natural Markovian reservoir with $\omega_c
\rightarrow \infty$. In this case the thermalization time is
simply inversely proportional to the coupling strength. For an
\lq\lq out of resonance\rq\rq~ engineered reservoir with $r \ll
1$, $\tau_T$ is notably increased and therefore the thermalization
process is slowed down.

As we will see in the following, there exist other two
characteristic timescales ruling the heating process: the period
of the system oscillator $\tau_S = 1/ \omega_0$ and the thermal
time $\tau_{\rm th}= 1/\nu_1= 1/ 2 \pi kT $ defined as the inverse
of the smallest positive Matsubara frequency. In Table
\ref{tab:times} we summarize the definitions of the four time
scales we have introduced up to now. In general the open system
dynamics depends strongly on the relative value of these four
characteristic time scales (see also \cite{haake2}).


Let us consider now the dynamics of the heating function for times
$ t \ll \tau_T$. For simplicity we consider as initial condition
the ground state of the system oscillator. The generalization to a
generic initial state is however direct and similar conclusions
hold. For times much smaller than the thermalization time,
Eq.~(\ref{hf}) can be approximated as follows:
\begin{eqnarray}
\langle n(t) \rangle &\simeq &  \int_0^t
\left(\Delta(t_1)-\gamma(t_1) \right) dt_1,  \label{hfgenericT}
\end{eqnarray}
where Eq.~(\ref{Gamma}) has been used. This equation shows that
the initial dynamics of the heating function depends strongly on
the sign of one of the time dependent coefficients of the secular
master equation (\ref{MERWA}). The reason for the heating function
to depend only on the coefficient $\Delta(t) - \gamma (t)$ and not
on $\Delta(t) + \gamma (t)$ is simply related to the initial
condition we have assumed. Indeed, when the initial state is the
ground state of the oscillator, for times $t \ll \tau_T$, the
probability of a jump up [absorbtion of a quantum of energy from
the reservoir, see Eq.(\ref{p1})] dominates over the probability
of a jump down [emission of a quantum of energy into the
reservoir, see Eq(\ref{p2})]. This second process, which is the
signature of the quantized nature of the reservoir, ensures the
thermalization.

\begin{table}[t]
\caption[t2]{\label{tab:times} Various time scales. }
\begin{tabular}{lcl}
\hline
         time scale name
       & symbol
       & explanation \\
\hline
      reservoir correlation& $\tau_R=1/\omega_c$ & $\omega_c=$ environment cutoff  \\
      thermalization &  $\tau_T=1/ \Gamma$  & $\Gamma=\alpha^2 \omega_0r^2 / (r^2 +1)$    \\
      system oscillator period & $\tau_S=1/\omega_0$ &  $\omega_0=$ oscillator frequency    \\
     thermal & $\tau_{\rm th}=1/\nu_1$ &$ \nu_1=$ Matsubara frequency   \\
reservoir memory & $\tau_{corr}=\tau_R$ &  for high T\\
                                & $\tau_{corr}=\tau_{th}$ &  for medium T \\
                               & $\tau_{corr}=\tau_{S}$ &  for low T \\
\hline
\end{tabular}
\end{table}

Equation (\ref{hfgenericT}) shows us that the coefficient
$\Delta(t)-\gamma(t)$ is the time derivative of the heating
function. Therefore, if $\Delta(t)
> \gamma(t)$ for all times $t \ll \tau_T$, the heating function grows monotonically, whereas if
there exist intervals of time in correspondence of which
$\Delta(t) < \gamma(t)$, the heating function decreases and
eventually oscillates. We remind that, for the case considered
here, whenever $\Delta(t) - \gamma(t) > 0$ at all times, the
master equation (\ref{MERWA}) is of Lindblad type, whilst if for
some time intervals $\Delta(t) - \gamma(t) < 0$ it is a
non-Lindblad type master equation.

\subsection{Lindblad and non Lindblad regions}
To better understand such a behavior we study in more details the
dynamics for three different regimes of the ratio $r$ between the
reservoir cutoff frequency and the system oscillator frequency: $r
\gg 1$, $r=1$ and $r\ll 1$. The first case corresponds to the
assumption commonly done when dealing with a natural reservoir
while the last case corresponds to an engineered \lq\lq out of
resonance\rq\rq~ reservoir.

By using Eqs. (\ref{gammasecord}) and (\ref{deltasecord}), a
straightforward but lengthy calculation  shows that, for $r \gg
1$, $\Delta (t) > \gamma(t)>0$. In this case the master equation
(\ref{MERWA}) is always of Lindblad type and the heating function
is a monotonically growing function. The three upper graphics of
Fig. \ref{results1} show the time evolution of the heating
function for $r=20$ in the case of low (a), intermediate (b) and
high (c) temperatures.

\begin{figure*}
\includegraphics[width=7.5 cm,height=5 cm]{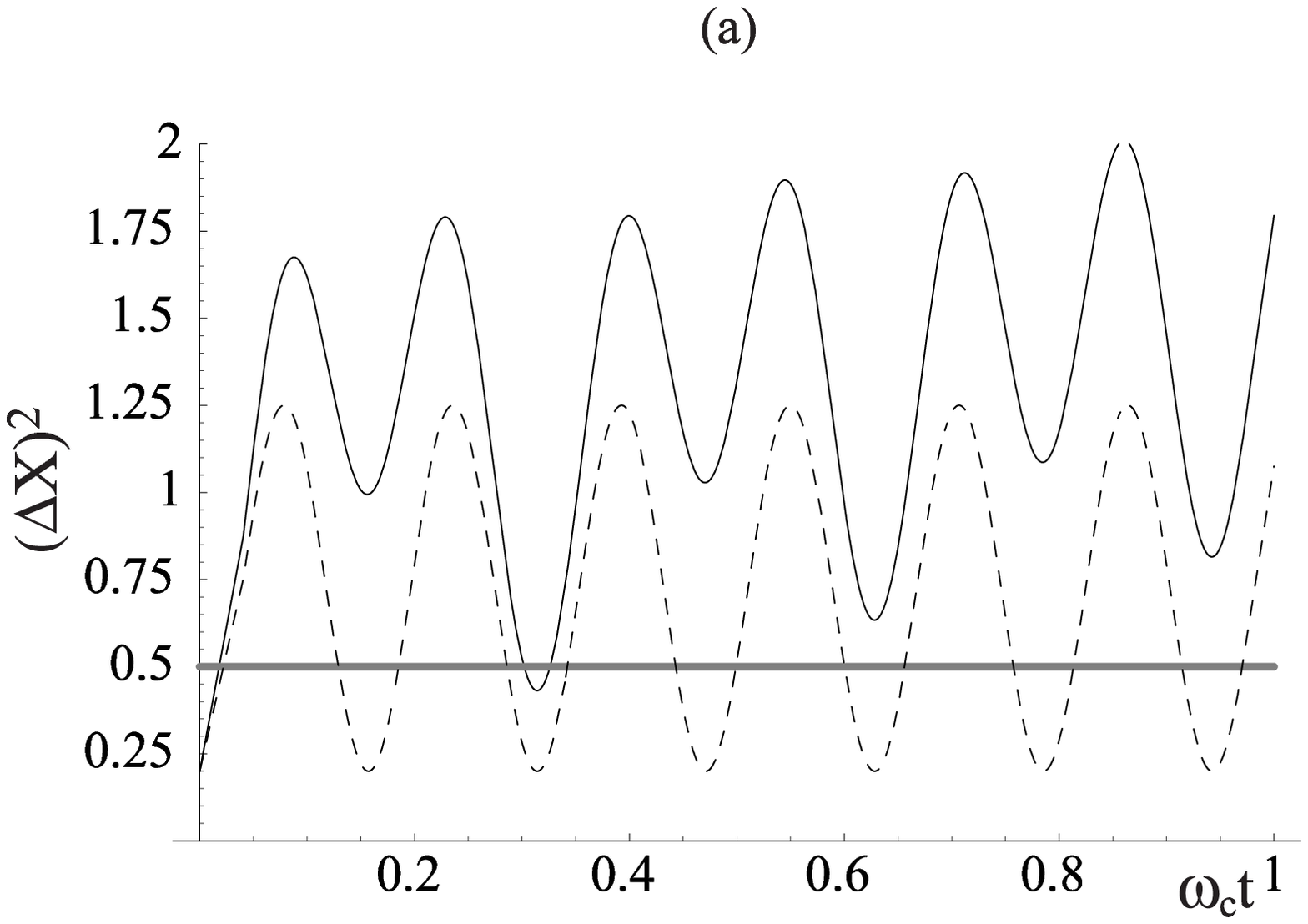}
\hspace{1cm}
\includegraphics[width=7 cm,height=5 cm]{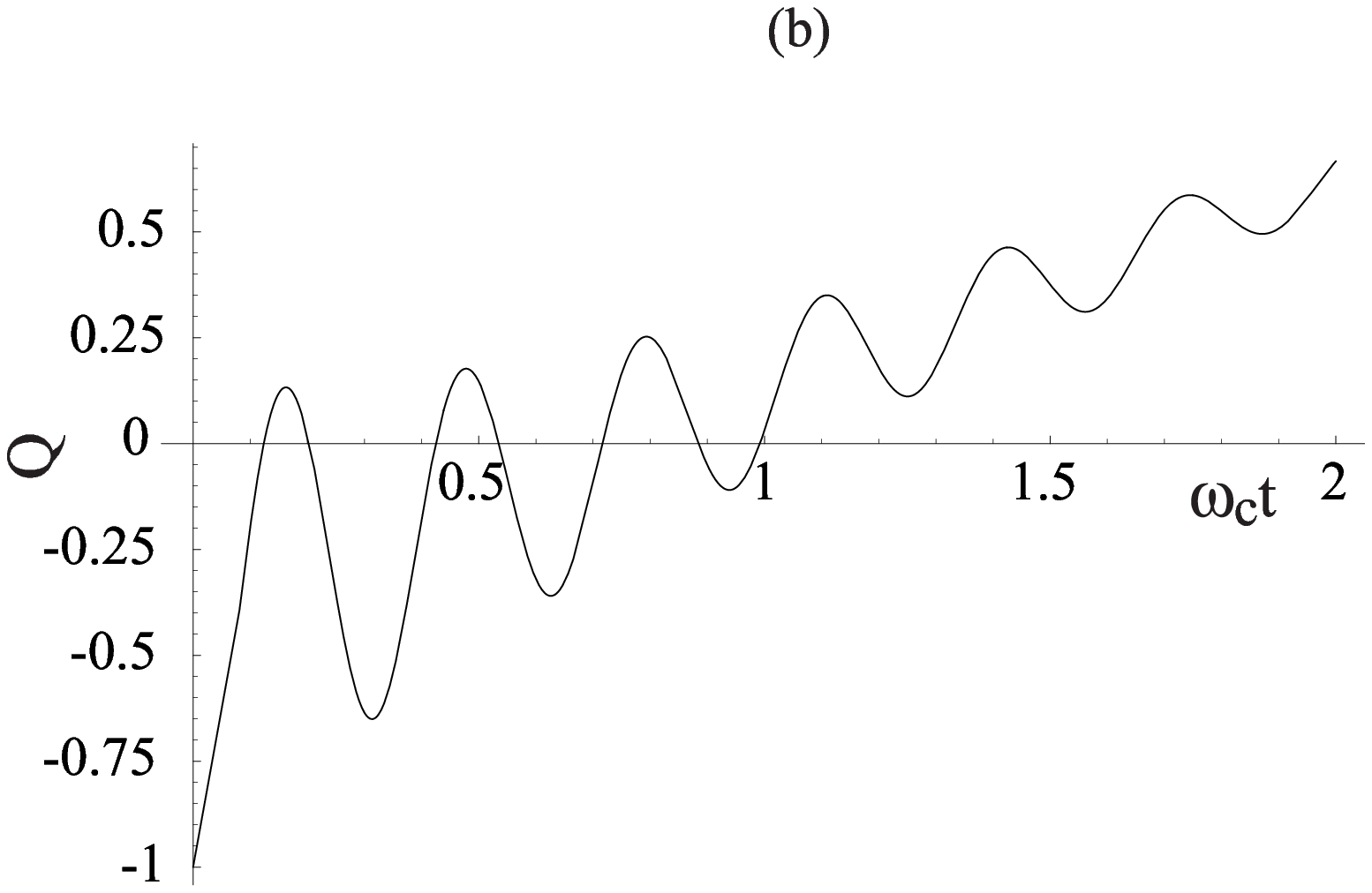}
\caption{\label{fig:squeezing} (a) Time evolution of the position
variance for an initial squeezed state having squeezing parameter
$s=0.4$. The reservoir parameters are $\alpha=10^{-2}$, $r=0.05$,
and $r_c = \omega_c /2 \pi KT = 0.2 \cdot 10^{-6}$. The dotted
line shows the variance of the free harmonic oscillator in absence
of coupling to the environment. The system exhibits squeezing
whenever $\left(\Delta X \right)^2(t)<0.5$. \\(b) Time evolution
of the Mandel parameter for an initial Fock state $\vert n =3
\rangle$. The reservoir parameters are the same as in (a)} .

\end{figure*}

In the case of an engineered \lq\lq out of resonance\rq\rq~
reservoir, that is when $r \ll 1$, the sign of the diffusion
coefficient is positive in the low temperature regime whilst for
intermediate and high temperatures it assumes negative values for
some time intervals. However, for intermediate and low
temperatures, there exist intervals of time in correspondence of
which $\Delta(t) - \gamma(t) < 0$. Whenever $\Delta(t)-\gamma(t)$
is negative, the heating function decreases, so the overall
heating process is characterized by oscillations as shown in
Fig.~\ref{results1} (g)-(h)-(i), where the dynamics of $\langle n
(t) \rangle$ for low, intermediate and high $T$, respectively, and
$r=0.1$ is plotted for $0< \omega_c t < 10$. The decrease in the
population of the ground state of the system oscillator, after an
initial increase due to the interaction with the reservoir, is due
to the emission and subsequent re-absorbtion of the same quantum
of energy. Such an event is possible since the reservoir
correlation time $\tau_R = 1/ \omega_c$ is now much longer than
the period of oscillation $\tau_s = 1/ \omega_0$. We underline
that, although the master equation in this case is not of Lindblad
type, it conserves the positivity of the reduced density matrix.
This of course does not contradict the Lindblad theorem since the
semigroup property is clearly violated for the reduced system
dynamics \cite{petruccionebook}.

Finally, for $r=1$, one can show numerically that $\Delta(t)
> 0$ at all times whatever is the reservoir temperature $T$.
Nonetheless, for intermediate and low temperatures the time
dependent coefficient $\Delta (t) - \gamma(t)$ assumes negative
values for some intervals of times. Such a situation leads again
to an oscillatory behavior of the heating function as shown of
Fig.~\ref{results1}(d) and Fig.~\ref{results1} (e).

It is worth stressing that the non-Markovian features of the
heating function discussed here do not depend on the secular
approximation. Indeed, as we have mentioned in Sec. II, Eq.
(\ref{hf}) coincides with the expression derived from the exact
solution. Stated in another way, the appearance of virtual
exchanges of energy between system and reservoir, characterizing
the non-Lindblad region, is a general feature of the non-Markovian
dynamics of the system and it is not connected with the secular
approximation.

\subsection{The key parameters $r$ and $T$}
The  border between the Lindblad and non-Lindblad regions depends
on two relevant reservoir parameters: its temperature $T$ and the
ratio $r$ between its cutoff frequency and the system oscillator
frequency. For high reservoir temperatures the quantity
effectively ruling the dynamics is the diffusion coefficient
$\Delta(t)$, since $\Delta(t) \gg \gamma(t)$. In other words, for
short times and high $T$, diffusion is always dominant with
respect to dissipation. Oscillatory dynamics of the heating
function appears for $r \ll 1$ since $\Delta(t)$ oscillates
assuming negative values. Figure~\ref{results2} (a) shows a
contour plot of $\Delta(r,t)$ for high $T$. The curve defined by
the equation $\Delta(r,t)=0$ for high $T$, with $\Delta(r,t)$
given by Eq.~(\ref{deltaHT}), defines the Lindblad non-Lindblad
border. From Fig.~\ref{results2} (a) one can see that the largest
value of $r$ in correspondence of which the system exhibits
non-Lindblad oscillatory heating is $r \simeq 0.27$.

For decreasing temperatures the amplitude of $\Delta(t)$ becomes
smaller and smaller. Thus, also in the presence of an oscillatory
behavior of $\Delta(t)$, that is when $r \ll 1$, for low
temperatures the diffusion coefficient remains always positive. In
this case, however, dissipation is not  negligible with respect to
diffusion anymore and their combined action is such that, for
intermediate and low temperatures, the non--Lindblad dynamics
appears already for $r > 1$ [see Fig.\ref{results2} (b)]. Stated
in another way, decreasing the temperature the oscillatory
behavior of the heating function appears for higher values of the
ratio $r$, which means that the non-Lindblad region becomes
larger. Fig.~\ref{results2} shows clearly that this region,
corresponding to negative values of $\Delta(t) - \gamma(t)$
($\Delta(t)$ for high $T$), is notably wider for low $T$ (b) than
for high $T$ (a).

While in this section we have investigated the border between the
Lindblad and non-Lindblad type regions, in the following section
we will concentrate on the non-Lindblad type dynamics for two
reasons. The first reason is that, in general, due to difficulties
in dealing with non-Lindblad type master equations, only  few
studies have been carried out in this regime. Secondly, we have
shown elsewhere that typical non-Lindblad dynamical features may
be experimentally revealed in the trapped ion context with
currently available technology \cite{Maniscalco04}.

\section{Non-Markovian dynamics of non-Lindblad type}\label{sec:results2}

In this section we focus on the dynamics of a quantum Brownian
particle in the case of interaction with an engineered out of
resonance reservoir, i.e.  for $r\ll 1$. By using the techniques
for reservoir engineering typical of trapped ion systems, such
region of the parameter space is already in the grasp of the
experimentalists. Indeed, by slightly modifying the experimental
conditions used in Ref. \cite{engineerNIST}, the oscillatory
non-Markovian dynamics of the heating function can be measured
\cite{Maniscalco04}.

\begin{figure*}
\includegraphics[width=14 cm,height=10 cm]{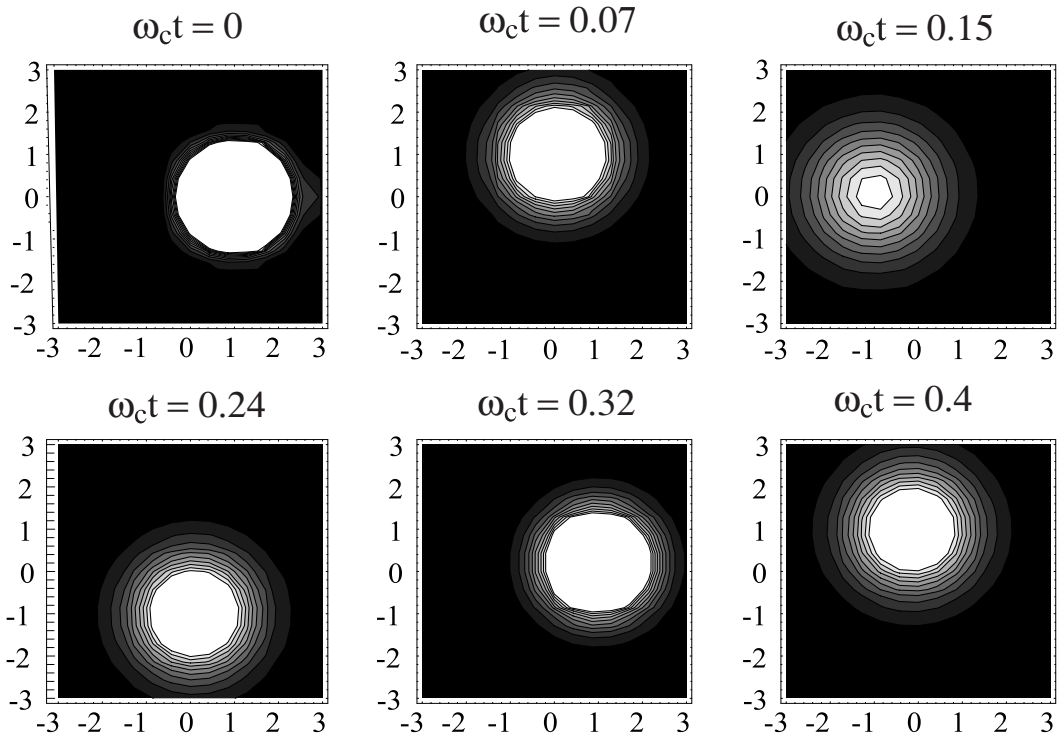}
\caption{\label{fig:wigner} Contour plots of the Wigner function
at different time instants. At $\tau=0$ the initial state is the
coherent state correspondent to $Re[\alpha_0]=1$,
$Im[\alpha_0]=0$. The reservoir parameters are $\alpha=10^{-2}$,
$r=0.05$, and $r_c= \omega_c /2 \pi KT =10^{-7}$.}
\end{figure*}

In order to characterize completely the dynamics of the quantum
Brownian particle in the non-Lindblad region we look at the
dynamics of the squeezing, of the Mandel parameter and of the
Wigner function for some exemplary initial states. We begin with
the analysis of the squeezing in position. By using Eq.
(\ref{xp}), it is possible to derive the following expression for
the variance of the dimensionless position operator $X$:
\begin{eqnarray}
\left( \Delta X \right)^2_t&=& e^{-\Gamma(t)} \left[\left( \Delta
X \right)^2_0 \cos^2(\omega_0 t) + \left( \Delta P \right)^2_0
\sin^2(\omega_0 t)  \right. \nonumber \\
&+& \left. C_0 \sin(2 \omega_0 t) \right] + \Delta_{\Gamma}(t)
\nonumber \\
&\equiv& e^{-\Gamma(t)} (\Delta X)_{\rm f}(t) + \Delta_{\Gamma}
(t) , \label{deltax2}
\end{eqnarray}
where $\left( \Delta X \right)^2_0$ and $\left( \Delta P
\right)^2_0$ are the initial variances of position and momentum
operators, respectively, and $C_0= \frac{1}{2}(\langle X_0 P_0 +
P_0 X_0 \rangle - \langle X_0 \rangle \langle P_0 \rangle)$ is the
initial position-momentum correlation function. In the last line
of Eq. (\ref{deltax2}) we have defined the function $(\Delta
X)_{\rm f}(t)$ describing the time evolution of the variance of a
quantum harmonic oscillator in absence of interaction with the
environment. In the high temperature limit, it is not difficult to
prove that the interaction with the environment generally causes
an increase in the variance of the position operator with respect
to its free dynamics. Indeed, for times much smaller than the
thermalization time $t \ll \tau_{T}$, Eq. (\ref{deltax2}) may be
approximated as follows
\begin{eqnarray}
\!\!\left( \Delta X \right)^2_t \!\!&\simeq& \!\!(\Delta X)_{\rm
f}(t) \!\!+ \!\! \!\int_0^t \!\!\! \Delta(t_1) - 2 (\Delta X)_{\rm
f}(t_1) \gamma(t_1), \label{deltax2a}
\end{eqnarray}
where Eqs. (\ref{Gamma})-(\ref{DeltaGamma}) have been used. Having
in mind that for high temperatures the condition $\Delta(t) \gg
\gamma(t)$ holds, one realizes that, provided that $(\Delta
X)_{free}(t)$ is not too big (corresponding to either a very high
initial squeezing in position/momentum or a very high initial
position-momentum correlation or a mixture of these cases), the
integral appearing in Eq. (\ref{deltax2a}) gives always a positive
contribution. In the short time non-Markovian regime, for low
reservoir temperatures, situations in which the system-environment
correlations lead to a decrease in the squeezing in position,
compared to its free time evolution, may in principle occur. Such
a situation would be of interest since it could be exploited to
generate squeezing through the interaction with an artificial low
temperature reservoir. We plan to investigate further this point
in the future.

In Fig. \ref{fig:squeezing} (a) we plot the short time behavior of
$\left( \Delta X \right)^2(t)$ for an initial squeezed state with
squeezing factor $s=0.4$, that is with $\left( \Delta X
\right)^2_0=s/2=0.2$ and $\left( \Delta P \right)^2_0=1/(2
s)=0.8$. We remind that $X$ and $P$ are dimensionless, therefore
squeezing in position corresponds to $\left( \Delta X
\right)^2<0.5$. In the figure we compare the time evolution of the
position variance for the damped harmonic oscillator with the case
of the isolated harmonic oscillator. From the figure one sees
clearly the effect of the virtual processes which tend to decrease
the position variance and to bring it back to its initial value.
The overall effect of the environment tends however to wash out
the initial squeezing.

Let us now consider the time evolution of the Mandel parameter
\cite{mandel}
\begin{equation}
Q(t)=\frac{\langle n^2 (t) \rangle - \langle n(t) \rangle
^2}{\langle n(t) \rangle}-1.
\end{equation}
This quantity gives an indication of the statistics of the
quantized mode described by the system oscillator. For a Fock
state  $Q$ takes its lowest value $Q=-1$ while  for a coherent
state $Q$ is equal to $0$ . Therefore, values of $Q<0$ indicate
subPoissonian statistics, while $Q = 0$ characterizes Poissonian
statistics and $Q>0$ superPoissonian statistics.  Using Eq.
(\ref{xp}) we have derived the time evolution of the Mandel
parameter as follows
\begin{eqnarray}
Q(t)\!=\!\frac{\langle n(t) \rangle^2 \!\!+ e^{-2 \Gamma(t)} \!
\langle n(0) \rangle \!\left[  Q(0) -\! \langle n(0)
\rangle\right]}{\langle n(t) \rangle}. \label{eq:Q}
\end{eqnarray}
In Fig. \ref{fig:squeezing} (b) we show the time evolution of the
Mandel parameter for an initial Fock state $\vert n = 3 \rangle$.
Due to the interaction with the artificial reservoir the initial
temporal evolution is characterized by oscillations between
subPossonian and Poissonian statistics of the quantized mode. This
behaviour may be traced back to the virtual photon exchanges
between the system and the reservoir and therefore is typical of
the non-Markovian non-Lindblad-type region. Looking at Eq.
(\ref{eq:Q}), and remembering that $\Gamma(t)>0$ , and that
$\langle n (t) \rangle \ge \langle n (0) \rangle$, it is easy to
convince oneselves that, if the initial state is Poissonian or
superPoissonian, i.e. $Q \ge 0$, the Mandel parameter will remain
positive at all times. In other words, the interaction with the
environment never creates subPoissonian statistics from an initial
Poissonian or superPoissonian statistics. This conclusion is valid
for generic temperatures $T$, provided that the weak coupling
assumption is satisfied.

We now look at the the system dynamics in the non-Lindblad regime,
 considering the time evolution of the Wigner function of an initial
coherent state $\vert \alpha_0 \rangle$. Having in mind Eq.
(\ref{chit}) and recalling that the Wigner function is simply the
Fourier transform of the quantum characteristic function,
\begin{equation}
W(\alpha) = \frac{1}{\pi^2} \int_{-\infty}^{\infty} d^2 \xi
\chi(\xi) e^{\alpha \xi^* - \alpha^* \xi},
\end{equation}
one gets
\begin{eqnarray}
W_t (\alpha) = \frac{1}{\pi \left[ \Delta_{\Gamma} (t) +
1/2\right]} \exp \left[ \frac{\left| \alpha_0 e^{-\Gamma(t)/2}
e^{-i \omega_0 t} - \alpha \right|^2}{\Delta_{\Gamma} (t) +
1/2}\right].
\end{eqnarray}
From this equation and from Fig. (\ref{fig:wigner}) one sees
clearly that the system-reservoir interaction  spreads the initial
Wigner function. Breathing of the Wigner function, that is the
oscillation in its spread, appears in correspondence of the
virtual processes. This is a new dynamical feature which is absent
both in the Markovian dynamics of the damped harmonic oscillator
and in the Lindblad-type non-Markovian dynamics. Indeed, in both
the previous regimes, the spread in the Wigner function simply
increases, linearly in time in the Markovian case, and
quadratically in time in the non-Markovian Lindblad-type case. We
note that different breathing scenarios for the second moments in
different regimes have been discussed in \cite{henkel}.

In summary, the exchanges of energy between system and reservoir
characterizing the non-Lindblad type region strongly influence the
dynamics of the system. In general, if the initial state of the
oscillator possesses nonclassical properties (as squeezing or
subPoissonian statistics), the interaction with the environment
tends to wash out such properties in a time scale which is
dependent, as one would guess, on the reservoir parameters
(spectral density and temperature). In this section we have
considered a high $T$ engineered \lq out of resonance\rq    (i.e.
with $r\ll 1$) reservoir. In this case the loss of nonclassical
properties appears in a time scale which is smaller or equal to
the reservoir correlation time $\tau_R=1/\omega_c$. Moreover, the
effect of the virtual processes, which is also important in this
time scale, may cause oscillations between classical and
nonclassical states, as in the case shown in Fig.
\ref{fig:squeezing} (b). It is worth noting that, in the situation
here considered, the loss of nonclassical properties, as well as
the oscillations due to virtual processes, happen in a time scale
which is in general much shorter than the decoherence time
$\tau_{\rm dec}= \lambda_T^2 / (d^2 \alpha^2)$, with $d$
separation between the two components of a quantum superposition
and $\lambda_T=\hbar/\sqrt{2 m k T}$ de Broglie wavelenght
\cite{zurekrev}. This is clearly related to the high temperature
condition of interest here,  which implies $\tau_R \gg 1/KT$.

In order to give a more quantitative estimate one should, however
look at some specific physical system, in order to fix also the
other quantities appearing in the definition of the two time
scales. To this aim we consider the recent experiment with trapped
ions in which the decoherence of different superpositions of the
vibrational motion of the center of mass of the ion was observed
\cite{engineerNIST}. For this system, this experiment has shown
that decoherence of a superpositon of two Fock states happens on a
time scale of the order of $30 \mu$s. Very recently we have
proposed an experiment \cite{Maniscalco04} to observe
non-Markovian features of the heating function in the same system
and with the same set up used in \cite{engineerNIST}. According to
our calculations, by slightly modifying the experimental
parameters used, one could observe oscillations in the heating
function in a time scale of the order of $3 \mu$s, that is one
order of magnitude less than the decoherence time measured in that
system for a superposition of Fock states.

In conclusion, the investigation carried out in this section sheds
light on the short time dynamics of the damped harmonic
oscillator, focussing in particular on the high $T$ regime.
Further analysis of the squeezing, of the Mandel parameter and of
the Wigner function may bring new insight in the time scales
governing the loss of nonclassical properties and in their
relationship with the decoherence and dissipation time scales. We
plan to explore further this  aspect, with particular attention to
the low temperature case.

\section{Conclusions}\label{sec:conclusions}
In this paper we have studied the short time non-Markovian
dynamics of a quantum Brownian particle moving in a harmonic
potential. The dynamics of this paradigmatic open quantum system
is described by a non-Markovian  master equation which is local in
time.  This master equation cannot be recast in the Lindblad form.
Nevertheless, under certain conditions, the master equation for
quantum Brownian motion is of Lindblad type, i.e. it has the same
operatorial form of the Lindblad master equation but with time
dependent (instead of constant) positive coefficients.

In the weak coupling limit, the relevant time dependent
coefficients can be cast in a closed form. In this case by using
the exact analytic solution in terms of the quantum characteristic
function, we have identified the parameters governing the passage
from Lindblad type to non-Lindblad type master equation. These
parameters are the reservoir temperature $T$ and the ratio $r$
between the frequency $\omega_0$ of the system oscillator and the
reservoir cutoff frequency $\omega_c$. It is worth stressing that
the weak coupling limit we consider in the paper is of interest
also in the light of the engineering of reservoir experiments. In
fact, in order to observe experimentally the key features of the
system-reservoir interaction, e.g. the role played by the
entanglement between system and reservoir in the decoherence
process, the coupling between system and reservoir does not have
to be too strong. The stronger is the coupling the faster is the
establishment of quantum correlations between the system and the
environment, and the more difficult is the experimental
observability of their dynamics. For this reason the techniques of
reservoir engineering, allowing to control both the coupling
constant and other reservoir parameters as its spectral density,
look very promising for investigating fundamental issues as the
quantum-classical border.

Our analysis of the short time non-Markovian region shows that the
Lindblad type dynamics is characterized by a monotone increase of
the heating function, and therefore of the energy, of the open
system. In the non-Lindblad type region, on the contrary,
oscillations in the mean energy of the system clearly indicate the
occurrence of virtual exchanges of energy between the system and
the reservoir. Lowering the reservoir temperature increases the
probability that virtual processes take place.

It is worth noting that whenever the master equation for the
system is of Lindblad type, it is possible to apply the standard
MC simulation schemes and there exists a direct correspondence
between the MC simulation method and a continuous measurement
scheme \cite{Dalibard92a,Molmer93a,Molmer96a}. For more general
non-Markovian Monte Carlo methods, e.g. the NMWF we have used in
this paper, an analogous correspondence would be of interest.
There are indications that the Lindblad non-Lindblad border might
be identified with the border between existence and non-existence
of a measurement scheme interpretation for non Markovian
stochastic methods. For this reason, the study of the dependence
of this border from parameters as the reservoir temperature and
the ratio $r$ might give some insight and useful hints to the
research on this contemporary topic.

Finally, last part of our paper deals with an analytic description
of the short time dynamics of the quantum Brownian particle when
virtual processes dominate. We have investigated in detail the
temporal evolution by looking at the squeezing properties, at the
Mandel $Q$ parameter and at the Wigner function. We have found
that, if the system initially possesses nonclassical properties as
squeezing in one of the quadratures or non-Poissonian statistics,
these properties tend to be washed out due to the interaction with
the reservoir. However oscillations between squeezing and
non-squeezing as well as between sub-Poissonian and Poissonian
statistics appears in connection with the virtual exchanges of
energy. A further sign of the virtual processes is the breathing
in the width of the Wigner function.

Summarizing, the main result of this paper is the detailed
analysis of the non-Markovian features characterizing the dynamics
of a quantum Brownian particle, with special attention to the
appearance of virtual processes for certain ranges of reservoir
temperature and cutoff frequency. Due to the generality of the
model here studied we think that our results can both contribute
to fundamental research on open quantum systems and, when applied
to specific physical contexts, shed light on their dynamics.

\section{Acknowledgements}
The authors gratefully acknowledge Heinz-Peter Breuer for helpful
comments and stimulating discussions. J.P. acknowledges financial
support from the Academy of Finland (projects 50314 and 204777),
and from the Magnus Ehrnrooth Foundation; and the Finnish IT
center for Science (CSC) for computer resources. J.P. and F.P.
thank the University of Palermo for the hospitality. J.P. and S.M.
acknowledge the EU network COCOMO (contract HPRN-CT-1999-00129)
for financial support. S.M. acknowledges the Angelo Della Riccia
Foundation for financial support and the Helsinki Institute of
Physics for the hospitality.

\appendix \label{appA}

\section{Diffusion coefficient}
In order to derive a closed analytic expression for $\Delta(t)$
valid for all temperatures and for all values of the ratio $r$ we
integrate Eq.~(\ref{kappa2}) and use the series expansion of the
hypergeometric function:
\begin{widetext}
\begin{eqnarray}
\Delta(t) &=& \alpha^2 \omega_0 \frac{r^2}{1+r^2} \Big\{ \coth
(\pi r_0) - \cot(\pi r_c) e^{-\omega_c t} \left( r \cos (\omega_0
t) - \sin
(\omega_0 t) \right) \nonumber \\
&+& \frac{1}{\pi r_0} \cos (\omega_0 t) \left[ \bar{F}(-r_c,t)+
\bar{F}(r_c,t) -
\bar{F}(ir_0,t) - \bar{F}(-ir_0,t) \right] \nonumber \\
&-& \frac{1}{\pi} \sin (\omega_0 t) \Big[ \frac{e^{- \nu_1 t}}{2
r_0(1+r_0^2) } \left( (r_0-i) \bar{G}(-r_0,t)+ (r_0+i)
\bar{G}(r_0,t) \right)  \nonumber \\
&+&  \frac{1}{2 r_c} \left( \bar{F}(-r_c,t) -
\bar{F}(r_c,t)\right) \Big] \Big\}. \label{deltasecord}
\end{eqnarray}
\end{widetext}
In this equation we have used the notations $r_0 = \omega_0/2 \pi
kT$, $r_c = \omega_c/2 \pi kT$,
\begin{eqnarray}
\bar{F}(x,t) &\equiv&   _2F_1 \left( x,1,1+x, e^{-\nu_1 t}
\right) \label{hyp1} \\
\bar{G}(x,t) &\equiv&   _2F_1 \left(2,1+x ,2+x, e^{-\nu_1 t}
\right),\label{hyp2}
\end{eqnarray}
where $_2F_1 \left( a,b,c,z \right)$ is the hypergeometric
function \cite{tavole}.

\section{Markovian and High Temperature limits} In the asymptotic
long time limit, the time dependent coefficients $\gamma(t)$ and
$\Delta(t)$ tend to their stationary value given by
\begin{eqnarray}
\Delta_M & \equiv & \Delta(t\rightarrow \infty) = \alpha^2
\omega_0 \frac{r^2}{1+r^2}  \coth (\omega_0/2 kT) \label{deltaM}
\\
\gamma_M & \equiv & \gamma(t\rightarrow \infty) = \alpha^2
\omega_0 \frac{r^2}{1+r^2}, \label{gammaM}
\end{eqnarray}
and the master equation (\ref{MERWA}) becomes the well known
Markovian master equation for a damped quantum harmonic oscillator
\begin{eqnarray}
\frac{ d \rho_S}{d t}= &-& \Gamma [n(\omega_0)+1] \left[ a^{\dag}
a \rho_S - 2 a \rho_S a^{\dag} + \rho_S a^{\dag} a \right]
\nonumber \\
&-& \Gamma n(\omega_0) \left[  a a^{\dag} \rho_S - 2 a^{\dag}
\rho_S a + \rho_S a a^{\dag}
 \right], \label{MarkovME}
\end{eqnarray}
with $\Gamma = \alpha^2 \omega_0 r^2/(1+r^2)$ and
$n(\omega_0)=\left( e^{\omega_0 / k T}-1\right)^{-1}$.

As far as the high temperature limit is concerned, having in mind
the definitions of $r_0$ and $r_c$, one sees immediately that such
an approximation amounts at taking $r_0, r_c \ll 1$, that is $x
\ll 1$. Under this condition one has \cite{tavole}:
\begin{eqnarray}
\bar{F}(x,t) &=&  _2F_1 \left( x,1,1+x, e^{-\nu_1 t} \right)
\nonumber \\
& \simeq&  _2F_1 \left( x,1,1, e^{-\nu_1 t} \right) = \left( 1-
e^{-\nu_1 t
}\right)^{-a}, \nonumber \\
\bar{G}(x,t) &=&  _2F_1 \left(2,1+x ,2+x, e^{-\nu_1 t} \right)
\nonumber \\
& \simeq&  _2F_1 \left(2,1 ,2, e^{-\nu_1 t} \right) = \left( 1-
e^{-\nu_1 t }\right)^{-1}. \nonumber
\end{eqnarray}
Inserting these expressions in Eq.~(\ref{deltasecord}) and using
the approximations
\begin{eqnarray}
\cot (\pi r_c) &\simeq& \frac{1}{\pi r_c} = \frac{2 kT}{\omega_c},
\nonumber \\
\coth (\pi r_0) &\simeq& 1+ \frac{1}{\pi r_0} \simeq \frac{2
kT}{\omega_0} \nonumber
\end{eqnarray}
one gets the Caldeira-Leggett high temperature expression for
$\Delta(t)$ \cite{Caldeira}
\begin{eqnarray}
\Delta(t)^{HT} &=& 2 \alpha^2 k T \frac{r^2}{1+r^2} \left\{ 1
- e^{-\omega_c t} \left[ \cos (\omega_0 t)\right.\right. \nonumber \\
&-&  (1/r) \left. \left. \sin (\omega_0 t )\right] \right\}.
\label{deltaHT}
\end{eqnarray}
The other time dependent coefficient, $\gamma(t)$, does not depend
on temperature as one can easily see by Eq.~(\ref{gamma}). We
stress that, comparing Eq.~(\ref{deltaHT}) with Eq.
(\ref{gammasecord}), one notices immediately that in the high
temperature regime, $\Delta(t) \gg \gamma(t)$.

\end{document}